# Effect of secondary swirl in supersonic gas and plasma flows in self-vacuuming vortex tube

VT Volov[1], AS Lyaskin[2]

[1]Department of Natural sciences, Samara State Transport University, Samara, Russia
[2]Department of Computational Fluid Dynamics, CADFEM-CIS, Samara, Russia

Email: vtvolov@mail.ru

**Abstract.** This article presents the results of simulation for a special type of vortex tubes – self-vacuuming vortex tube (SVVT), for which extreme values of temperature separation and vacuum are realized. The main results of this study are the flow structure in the SVVT and energy loss estimations on oblique shock waves, gas friction, instant expansion and organization of vortex bundles in SVVT. Keywords: self-vacuuming vortex tube, supersonic swirling gas flows, oblique shock wave, rotating cord.

**Nomenclature**
$t$ – time, $x_i$ – Cartesian coordinates, $u_i$ – Reynolds-averaged components of instantaneous velocity vector in the Cartesian coordinate system, $u_i'$ – pulsation components of instantaneous velocity vector in the Cartesian coordinate system, $\rho$ – density, $p$ – pressure, $h_t$ – total enthalpy, $T$ – temperature, $\lambda$ – coefficient of thermal conductivity, $C_p$ – specific heat capacity at constant pressure, $\mu$ – dynamic viscosity, $\overline{\rho u_i' u_j'}$ – components of turbulent stress tensor, $k$ – specific turbulent kinetic energy, $\omega$ – specific rate of dissipation of $k$, $P^*$ - total (stagnation) pressure, $T^*$ - total (stagnation) temperature.

## 1. Introduction
Vortex flows have been of significant interest since the mid-20th century because of their occurrence in industrial applications such as furnaces, gas turbines and collectors (Gupta et al. [1]). Besides, vortex (or high degree of swirling) can also produce a hot and cold stream separation via Ranque-Hilsch vortex tube [2, 3]. Nash [4] and Dobratz [5] provided extensive reviews of vortex tube utilizations and enhancements. Intense experimental and numerical studies of Ranque-Hilsch tubes have begun since 1931 and continue even nowadays [6, 7, 8]. The most famous scientific school studying the Ranque-Hilsch effect in Russia was led by Professor A.P. Merkulov [9, 10, 11]. In addition to traditional applications of vortex tubes, this school proposed for them a number of new utilizations in aviation, chemical industry, power industry, agriculture, etc. Special merit of this scientific school is the creation of vortex electro-discharged devices: $CO_2$-lasers and plasmatrons [12, 13]. A significant number of theories have been proposed to explain the vortex effect of «temperature separation» since its initial observations by Ranque [2]. Despite all proposed theories, none has been able to fully explain the one. This fact is partially compensated by the numerical models of flow in vortex tubes. Numerical studies of swirling gas flows in the vortex tube were presented by following researches [6, 7, 8]. Simulations were made using turbulent models ranging from simple RANS (as standard *k*-ε) [6] to large eddy simulation (LES) [7]. At the same time the least developed type of vortex tubes in theoretical and experimental aspects is the self-vacuuming vortex tube (SVVT), where extreme values of temperature separation ($\Delta T_{cold}$=152K, gas - air [1]) and vacuum ($\pi^* = P^*/P_{axis}$=40, where $P^*$- total gas pressure at the inlet of SVVT, $P_{axis}$- gas pressure near the axis of SVVT) are be realized. In contrast with classic vortex tubes [2, 3] SVVT do not have hot and cold flow exits. Instead, SVVT is used for cooling of its internal body. There are some engineering models for calculation of flows and thermodynamic performances of SVVT [13, 14]. Though they allowed creating a new class of electro-discharged devices ($CO_2$-lasers and plasmatrons [12]), still there are no theoretical studies of the flow behaviour in SVVT based on modern computational fluid dynamics (CFD) simulation technics. The present paper is devoted to the CFD study of strongly swirling gas flows in SVVT based on the RSM turbulence model.

## 2. Schematics of the problem geometry
Schematics of the problem geometry is shown at figure 1. The basic dimensions of SVVT are the following: tube's radius and height are $R_t$ = 5 *mm*, $H_t$ = 11.5 *mm*, respectively. Central body radius, fillet radius, diffusor radius and diffusor height are $R_{cb}$ = 0.5 *mm*, $R_f$ = 1.5 *mm*, $R_d$ = 25 *mm*, $H_d$ = 1 *mm*, respectively. The width of inlet is *a* = 2.1 *mm* and its height is *b* = 1.4 *mm*.

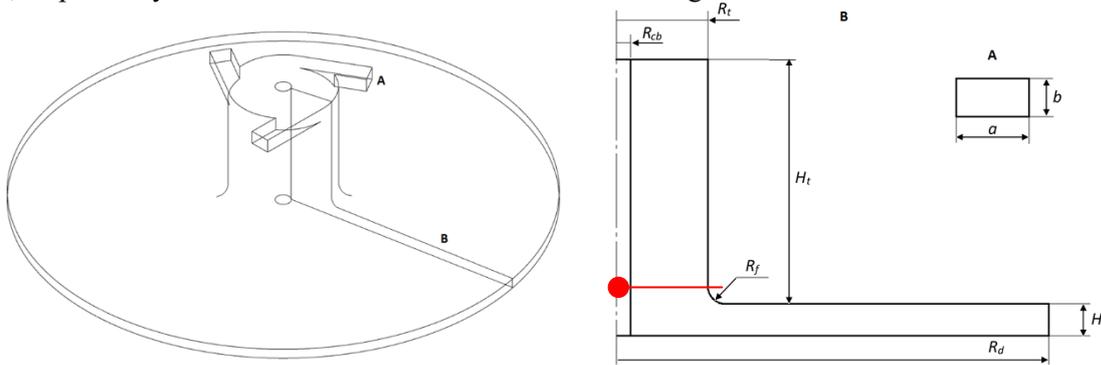

Figure 1. Schematics of the problem geometry (red dot marks reference frame origin)

## 3. Mathematical model
The compressible turbulent flows in SVVT are governed by mass, momentum and energy conservation equations, together with equation of state. The Reynolds-averaged forms of these equations are:

$$\frac{\partial \rho}{\partial t} + \frac{\partial (\rho u_k)}{\partial x_k} = 0,$$

$$\frac{\partial (\rho u_i)}{\partial t} + \frac{\partial (\rho u_i u_k)}{\partial x_k} = -\frac{\partial p}{\partial x_i} + \frac{\partial}{\partial x_k}(\tau_{ki}^{eff}),$$

$$\frac{\partial (\rho h_t)}{\partial t} - \frac{\partial p}{\partial t} + \frac{\partial (\rho u_k h_t)}{\partial x_k} = \frac{\partial}{\partial x_k}\left[\lambda\left(\frac{\partial T}{\partial x_k}\right)\right] + \frac{\partial}{\partial x_k}(u_i \tau_{ki}^{eff}),$$

$$\rho = \frac{Wp}{R_0 T},$$

where $h_t = h + \frac{1}{2} u_k u_k$ is total enthalpy, $h = C_p T$ is static enthalpy, $\tau_{ij}^{eff} = \tau_{ij} - \rho \overline{u_i' u_j'}$ is the effective stress tensor, $\tau_{ij} = \mu\left(\frac{\partial u_i}{\partial x_j} + \frac{\partial u_j}{\partial x_i}\right) - \frac{2}{3}\frac{\partial u_i}{\partial x_j}\delta_{ij}$ is the viscous stress tensor, $W$ is molar mass, $R_0 = 8.314472 \ m^2 \cdot kg/(s^2 \cdot K \cdot mol)$ is the universal gas constant.

In this study, the flow was considered fully turbulent. For the description of turbulence, Omega Reynolds Stress model was used [14]. It comprises six transport equations for turbulent stresses as well as an additional transport equation for ω, which is similar to the equation of *k*-ω model.

## 4. Numerical model

For the numerical solution of governing equations, ANSYS CFX commercial CFD software was used. It utilizes a finite volume method. The simulation was carried out with the following boundary conditions: specified values of total (stagnation) pressure and total (stagnation) temperature at inlets, with flow direction at inlets normal to the boundaries; specified values of pressure and temperature at the exit. These values were interpreted as values of static parameters at those parts of boundary where the flow leaves the simulation domain and as values of total parameters at those parts of boundary where flow enters the simulation domain in case of reversed flow. All the walls of SVVT were considered adiabatic. The inlet parameters varied: $P^*$ from 0.15 *MPa* to 0.7 *MPa*, $T^*$ from 500*K* to 10,000*K*. The gas was helium, with $W = 4.0$ *g/mol*, $C_p = 5193.0$ *J/(kg·K)*. Transport properties λ and μ were considered as functions of temperature according to [15] for $T^*<5000K$ and calculated from kinetic theory for $T^*\geq5000K$.

The calculation was carried out in a pseudo-transient formulation with fixed pseudo-time step equal to $10^{-7}$ s. Convergence criteria, besides the standard ones based on average normalised residuals, were the global imbalances of mass and energy (the threshold value 0.1%) and constancy of mass flow rate.

The initial computational mesh featured $4.3\cdot10^6$ elements ($1.5\cdot10^6$ nodes) with the following adaptation, based on pressure gradient, up to $11.3\cdot10^6$ elements ($3.0\cdot10^6$ nodes).

## 5. Results and discussion

Comparison of viscous and inviscid flow simulation results revealed that while direct friction and turbulent losses are rather small (from 1 to 10% of power lost due to friction and less than 1% of power lost due to turbulence), viscous effects play a significant role, drastically changing the general flow pattern. Viscous effects are responsible for interaction of "streams" produced by separate inlets. This interaction twists these "streams", turning them into rotating cords (secondary vortices) with high intrinsic swirl (see Figures 2 and 3; results obtained for $T^* = 500K$, $P^* = 0.3MPa$). Such "flow rearrangement" could be considered as a source of energy losses.

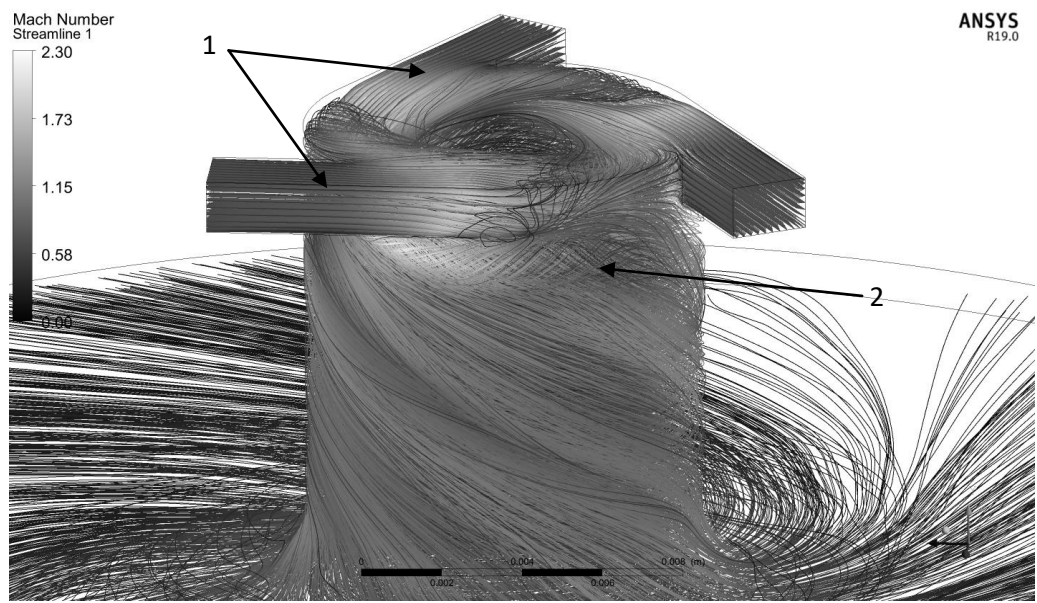

Figure 2: Streamlines originating from SVVT inlets colored by local Mach number: 1 – unswirled initial "streams", 2 – rotating cord (secondary vortex) resulting from their interaction.

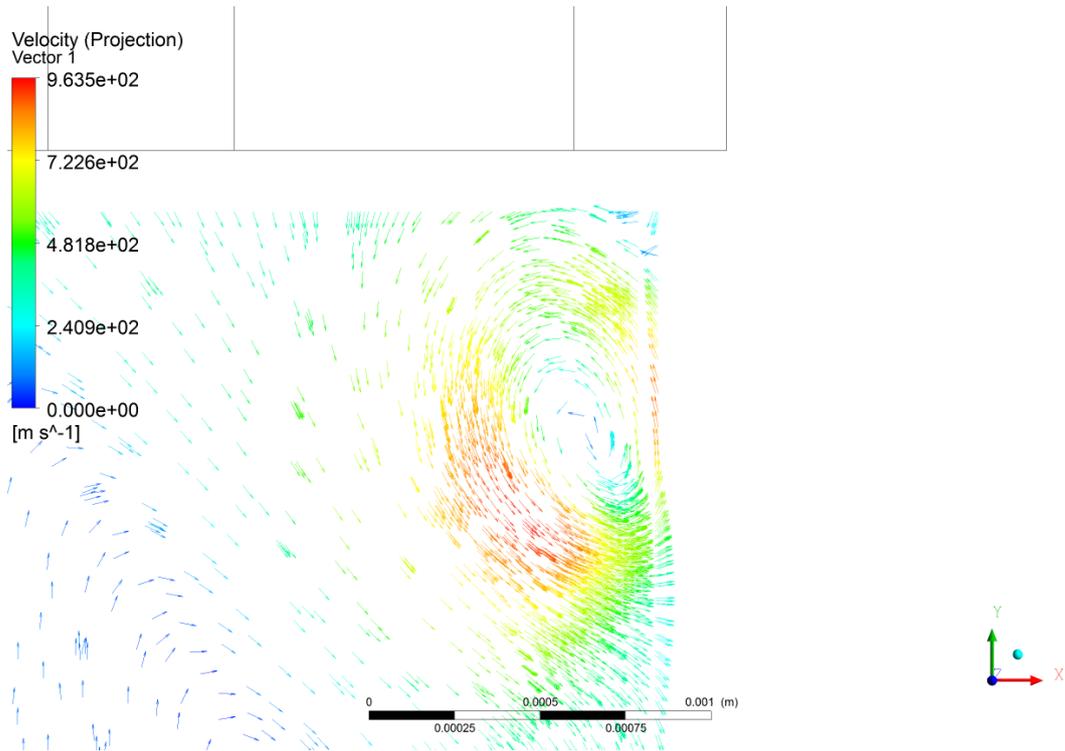

Figure 3: Cross-section of one of rotating cords formed from inlet "streams" interaction: velocity vectors projected to section plane.

As "streams" become supersonic due to expansion from inlet channels into main chamber, their interaction with each other and with SVVT peripheral wall also produces a system of oblique compression shocks. These shocks could be considered as another source of energy losses. Note that viscous effects play a significant role here, too – the pattern of shocks is very different between inviscid and viscous flow (see Figure 4).

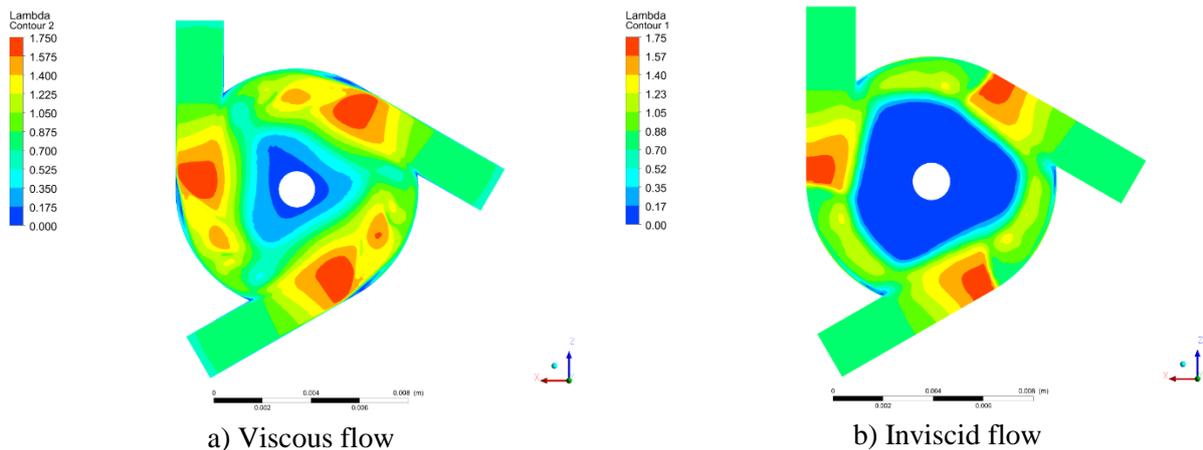

a) Viscous flow    b) Inviscid flow

Figure 4: Velocity coefficient in the inlet plane for viscous and inviscid flow.

Interestingly, it was found that for lower end of the considered temperature range (500 – 10,000K) the main flow could remain supersonic even after passing the shocks (see Figure 5). This was because addition of transonic "relative" velocity (associated with high intrinsic swirl of rotating cords produced by "streams" interaction) to transonic circumferential velocity (associated with general rotation, produced by tangential gas supply) results in supersonic total velocity.

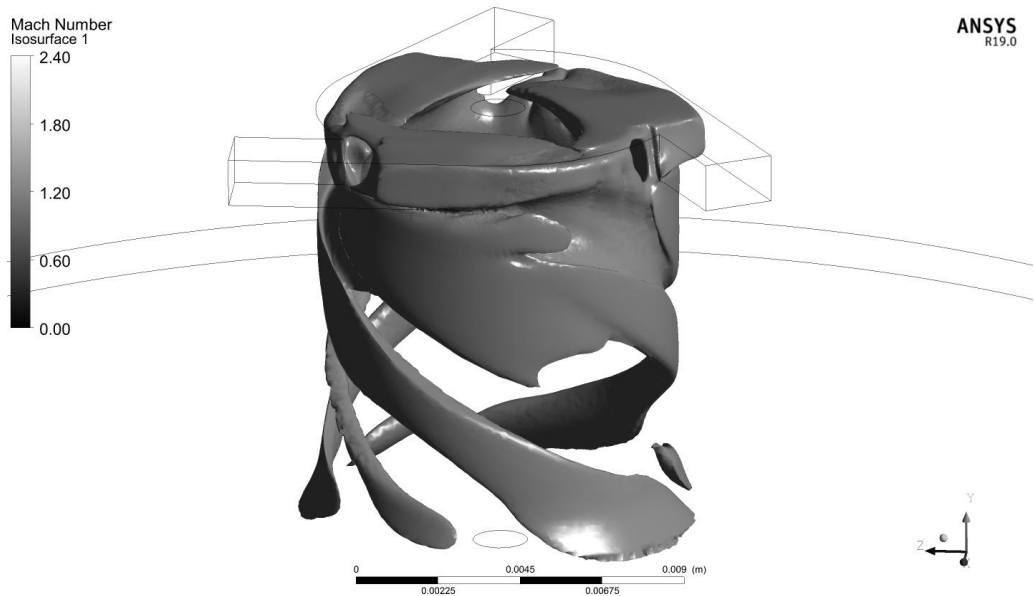

Figure 5: Isosurface of Mach number equal to 1.

While qualitative analysis of energy losses due to compression shocks presented mere technical difficulties, for qualitative analysis of energy losses due to "flow rearrangement" and swirl generation it was necessary to estimate the energy associated with the intrinsic swirl of rotating cords. Taking the curl of velocity vector

$\boldsymbol{\omega} = \nabla \times \mathbf{V}$,

and considering it in cylindrical reference frame aligned with SVVT main chamber (i.e. axial coordinate $z$ corresponding to Y coordinate of global Cartesian reference frame at Figures 1 and 2) it could be assumed that: its axial component $\omega_z$ could be attributed to "general" or "main vortex" rotation, generated by tangential gas supply; its circumferential component $\omega_\theta$ could be roughly attributed to "secondary" rotation or intrinsic swirl of rotating cords. By analogy with enstrophy, it was proposed to integrate $\omega_\theta^2$ by surface at section plane $\theta = const$ to get a crude (by the order of magnitude) estimation of specific energy associated with swirl (note that such a section plane must be truncated in order to omit boundary layer, containing a lot of vorticity). For more precise estimation it was necessary to identify the axis of the rotating cord (to construct the section plane normal to it) together with its bounds and consider the curl of velocity in a local coordinated frame aligned with rotating cord axis.

## 6. Conclusion

For the first time flow pattern in SVVT was studied with state-of-the-art CFD technics. It was found that this pattern is rather complex: it features both subsonic and supersonic regions, systems of oblique shocks, primary rotation (main vortex) and "rotating cords" (secondary vortices) with strong interaction. An attempt was made to estimate the sources of power losses. It was found that only a small amount of losses could be attributed to friction and turbulence (from 2% to 10% due to friction, depending on inlet conditions, with losses due to turbulence making less than 1% for all cases). It was assumed that most of the losses occur due to shocks and secondary swirl generation ("rotating cords" formation).